\documentclass[twocolumn,showpacs,preprintnumbers,amsmath,amssymb,amsfonts,superscriptaddress,floatfix,prl,aps]{revtex4-2}

\usepackage[paperwidth=210mm,paperheight=297mm,centering,hmargin=2cm,tmargin=1.6cm,bmargin=3.cm]{geometry}

\usepackage{booktabs}
\usepackage{subfigure}
\usepackage{graphicx}
\usepackage{amsfonts}
\usepackage{amssymb}
\usepackage{amsmath}
\usepackage{overpic}
\usepackage{siunitx}
\usepackage{braket}
\usepackage{xcolor}
\usepackage{cancel}

\usepackage[utf8]{inputenc}
\usepackage{lipsum}
\usepackage[T1]{fontenc}
\usepackage[newcommands]{ragged2e}
\usepackage[font=small,labelfont=bf,justification=Justified]{caption}
\usepackage{xcolor}
\usepackage[english]{babel}

\newcommand{\beq}{\begin{equation}}
\newcommand{\eeq}{\end{equation}}
\newcommand{\bc}{\begin{center}}
\newcommand{\ec}{\end{center}}
\newcommand{\bee}{\begin{eqnarray}}
\newcommand{\eee}{\end{eqnarray}}

\DeclareUnicodeCharacter{2212}{-}

\makeatletter
\makeatletter \renewcommand{\fnum@figure}{{\bf{\figurename~\thefigure}}}

\makeatother

\begin{document}

\title{ Revisiting vestigial order in nematic superconductors: gauge-field mechanisms and model constraints}

\author{I. Maccari} 
\email{imaccari@phys.ethz.ch}
\affiliation{Institute for Theoretical Physics, ETH Zurich, CH-8093 Zurich, Switzerland}
\author{E. Babaev}
\affiliation{Department of   Physics, The Royal Institute of Technology, Stockholm SE-10691, Sweden}
\affiliation{Wallenberg Initiative Materials Science for Sustainability, Department of Physics, KTH Royal Institute of Technology, SE-106 91 Stockholm, Sweden}
\author{J. Carlström}
\affiliation{Department of   Physics, The Royal Institute of Technology, Stockholm SE-10691, Sweden}
\affiliation{Department of  Physics, Stockholm University, Stockholm SE-10691, Sweden}

\date{\today}

\begin{abstract}

{An electronic nematic order that originates from superconducting fluctuation but persists above the superconducting transition temperature is often referred to as a vestigial nematic phase. Such a vestigial order
belongs to the broader class of composite orders discussed in earlier literature, characterized by ordering
in gauge-invariant combinations of superconducting order parameters while the individual superconducting
order parameters remain disordered. These states include metallic superfluids, paired phases, and composite
(charge-4e) superconductors. Whether and under what conditions such a vestigial phase can emerge in realistic
models of nematic superconductors remains an open question. Recent analytical work [P. T. How and S. K.
Yip, Phys. Rev. B 107, 104514 (2023)] concluded that vestigial nematic phases—and related mechanisms—do
not appear in the widely studied models proposed for, e.g., Bi$_2$Se$_3$-based candidates. 
 To shed light on this question, we perform large-scale Monte Carlo simulations of a three-dimensional Ginzburg-Landau model of
a nematic superconductor. Consistent with the findings of How and Yip, our numerical results confirm that commonly considered models do not exhibit vestigial nematic phases or nematic-fluctuation-induced charge-4e superconductivity. Extending the analysis to include coupling to a gauge field, we show that vestigial nematic
order can, under restrictive conditions, be stabilized through an alternative mechanism: intercomponent coupling
mediated by the gauge field or the effects of strong correlations.}

\end{abstract}

\maketitle

\section{Introduction}
Significant experimental and theoretical efforts are currently directed toward identifying states characterized by order parameters composed of four fermionic fields, understanding the conditions under which such states emerge, and identifying possible candidate materials.  
These {composite orders} can emerge in systems with a superconducting ground state, resulting in an order parameter formed by  four--rather than two--electrons. While standard Bardeen–Cooper–Schrieffer (BCS) theory forbids their formation, such states can arise when two key assumptions of BCS theory are violated:
(i) the superconducting order parameter breaks multiple symmetries, and
(ii) strong fluctuations invalidate the BCS mean-field approximation, producing a regime of incoherent Cooper pairs, {i.e. no order in bilinear fields, which still
preserve higher-than-bilinear broken composite symmetries. Thus, in this regime, the only symmetries spontaneously broken are composite symmetries constructed from quartic or higher-order electronic operators.}

The terminology in this field is not yet fully settled, and similar types of order have been discussed earlier under different names, depending on the  context. In the recent literature on high-$T_c$ superconductors, such phases are often referred to as \emph{vestigial order}~\cite{fradkin_colloquium_2015}. Other terms that appear are \emph{electron quadrupling condensates} (particularly in connection with charge-$4e$ and counterflow condensates), \emph{quadrupling phases}, and \emph{composite order}--although the latter is also used in other contexts with a different meaning. In related but distinct microscopic settings, analogous phases have been labeled \emph{symmetric mass generation}~\cite{butt2021symmetric} and \emph{paired phases}~\cite{kuklov_deconfined_2006,kuklov_deconfined_2008}.

{
In this work, we adopt the term composite order--as used in earlier studies--and employ it synonymously with vestigial order used for the same phenomena in several recent works on superconducting systems. The stability of such phases is highly sensitive to the symmetries that are broken and to the spatial dimensionality of the model. While some progress can be achieved using analytical arguments, different mechanisms have been proposed for their stabilization, including gauge-field–mediated intercomponent coupling ~\cite{babaev2002phase,babaev_superconductor_2004} and partial melting of pair-density-wave order \cite{agterberg_dislocations_2008, radzihovsky_quantum_2009,gaggioli2025spontaneous,agterberg_conventional_2011}. However, establishing their existence in three-dimensional superconducting models remains challenging. Indeed, recent works ~\cite{how_absence_2023,yuan_absence_2024, How_Superfluid2024,how_broken_2024} have demonstrated that simplified analytical approaches may lead to false positives, predicting composite order in systems where it does not actually occur. This underscores the need for large-scale Monte Carlo simulations, which fully capture superconducting and nematic fluctuations—including topological excitations and their interactions—and are therefore typically required to reliably determine the phase diagram.}

 
The emergence of fermion quadrupling condensates in two- and three--dimensional London and Ginzburg--Landau models has been demonstrated numerically in several classes of superconductors with broken symmetries. These include
 (i)
  $U(1)\times U(1)\to U(1)$ \cite{ smorgrav_observation_2005,kuklov_deconfined_2006, herland_phase_2010},  
(ii) $U(1)\times Z_2 \to Z_2$ as in $s+is$, $s+id$, $d+id$, and $p+ip$ superconductors \cite{bojesen_time_2013,maccari_prediction_2023, how_broken_2024,bojesen_phase_2014, grinenko_state_2021, maccari_effects_2022},  
(iii)$SU(2)\to O(3)$ \cite{kuklov_deconfined_2008, motrunich_comparative_2008,herland_phase_2013}
(iv) $SU(N)\to SU_n(N)$ \cite{weston_composite_2021,bonati2025three,bonati2021lattice,bonati2023abelian,bonati2023coulomb}.

Here, the remaining broken symmetries $U(1),Z_2,O(3)$ are associated with a four-electron composite order.
{The trend observed is that it is typically harder to stabilize these phases when associated to   higher broken symmetries.}
  
Analogous bosonic counterparts in $U(1)\times U(1) \to U(1)$ and higher symmetries, stabilized by strong interaction effects, have also been proposed and demonstrated via Monte Carlo simulations at the level of both effective field theories and fully microscopic quantum models \cite{kuklov_counterflow_2003, altman_phase_2003, kuklov_deconfined_2006, kuklov_superfluid-superfluid_2004, soyler_sign-alternating_2009, sellin_superfluid_2018, blomquist_borromean_2021,Hubener2009,Powell2009,Hu2009,Menotti2010,Hu2011,Schachenmayer2015,Venegas_Gomez2020B,deParny2021,Basak2021,kuklov2025field}, and are
a subject of ongoing experimental research \cite{zheng_counterflow_2025}. Related composite orders were also studied  in high-energy physics models with various symmetries using large-scale numerical studies \cite{bonati2024diverse, bonati2023abelian, bonati2024deconfinement, butt_symmetric_2025,bonati2025three}.

{Overall composite orders in superconductors have been studied, especially in conjunction with $U(1)\times Z_2 \to Z_2$ phase transitions.
Experimentally, 
a $U(1)\times Z_2 \to Z_2$ phase transition scenario has been investigated in Ba$_x$Fe$_{1-x}$Fe$_2$As$_2$ \cite{grinenko_state_2021, shipulin_calorimetric_2023, halcrow_probing_2024, bartl2025evidence}, which, at low temperatures, realizes a $U(1)\times Z_2$ superconducting state that breaks time-reversal symmetry. 
Above the superconducting transition, the system retains long-range order associated with the phase difference between pairs of preformed Cooper pairs in different bands, thereby realizing a $Z_2$ electron quadrupling condensate with a rich array of novel properties \cite{grinenko_state_2021, shipulin_calorimetric_2023, halcrow_probing_2024, bartl2025evidence}.
{For different symmetries,  }realization of a related bosonic system with ultra-cold atoms in an optical lattice was recently reported in~\cite{zheng_counterflow_2025}.
{Ongoing research also pursues the realization of charge-4e superconductivity \cite{ge_charge-4e_2024}.}

This work focuses on a class of systems that is less studied numerically: nematic superconductors that break a $U(1) \times Z_3$ symmetry. Here, $Z_3$ is associated with the space-rotational symmetry. For derivations of the corresponding mean-field models, see~\cite{fu_odd-parity_2010, fu_oddnematic_2014,zyuzin_nematic_2017}.
Popular candidate materials for nematic superconductivity are Bi$_2$Se$_3$-based materials ~\cite{matano_spin-rotation_2016, pan_rotational_2016, venderbos_identification_2016, asaba_rotational_2017, yonezawa_thermodynamic_2017, shen_nematic_2017, smylie_superconducting_2018, yonezawa_nematic_2019, how_signatures_2019}.
Remarkably, an experimental study \cite{cho2020z} claimed signatures of nematicity that persist above the superconducting transition temperature. The observations were interpreted as evidence for a $Z_3$ vestigial nematic phase—i.e., rotational symmetry breaking driven by non-condensed electron pairs.
The authors of~\cite{cho2020z}  made a careful reservation that their sample does not, strictly speaking, possess $Z_3$ symmetry--since only one of the $Z_3$ states is realized upon repeated cool downs. Nevertheless, their study sparked new interest in the theoretical underpinnings of $U(1) \times Z_3$ superconductors.
{Nematicity arising slightly above the superconducting phase transition was also reported in the experiment \cite{ddvn-8c9n}.}

{The interpretation in \cite{cho2020z} was based on the theoretical framework proposed in Ref.~\cite{hecker_vestigial_2018}, which relies on an effective model for the nematic order obtained by introducing auxiliary bilinear nematic fields and treating superconducting fluctuations at the Gaussian (one-loop) level. More recently, Ref.~\cite{how_absence_2023} reached an opposite conclusion. By computing bilinear susceptibilities directly from the weak-coupling Ginzburg-Landau propagators, the authors concluded that these nematic models~\cite{hecker_vestigial_2018, fernandes2019intertwined}   do not support vestigial phases.}
While this work represents the most analytically advanced treatment of the problem to date, it highlights two important directions for further investigation:
(i) large-scale Monte Carlo simulations of the proposed model{, to fully account for all superconducting and nematic fluctuations, including topological excitations and their mutual interaction,} and  
(ii) exploration of whether the composite nematic order in superconductors can occur via alternative mechanisms involving Cooper pairing.

In this work, we report large-scale Monte Carlo simulations of a three-dimensional model with a nematic superconducting ground state.  
Our result is {two-fold. First, we} show that, consistent with the conclusions of~\cite{how_absence_2023}, the basic models under consideration do not exhibit nematic order above the superconducting critical temperature.
{Second, we extend the model to include a finite coupling to a gauge field and explore an alternative mechanism whereby composite order (vestigial nematicity) emerges through gauge-field–mediated intercomponent coupling.
We show that this mechanism can give rise to a composite nematic phase. However, in three dimensions, the realization of this phase requires a very strong gauge coupling. }

\begin{figure}[t!]
    \centering
    \includegraphics[width=0.85\linewidth]{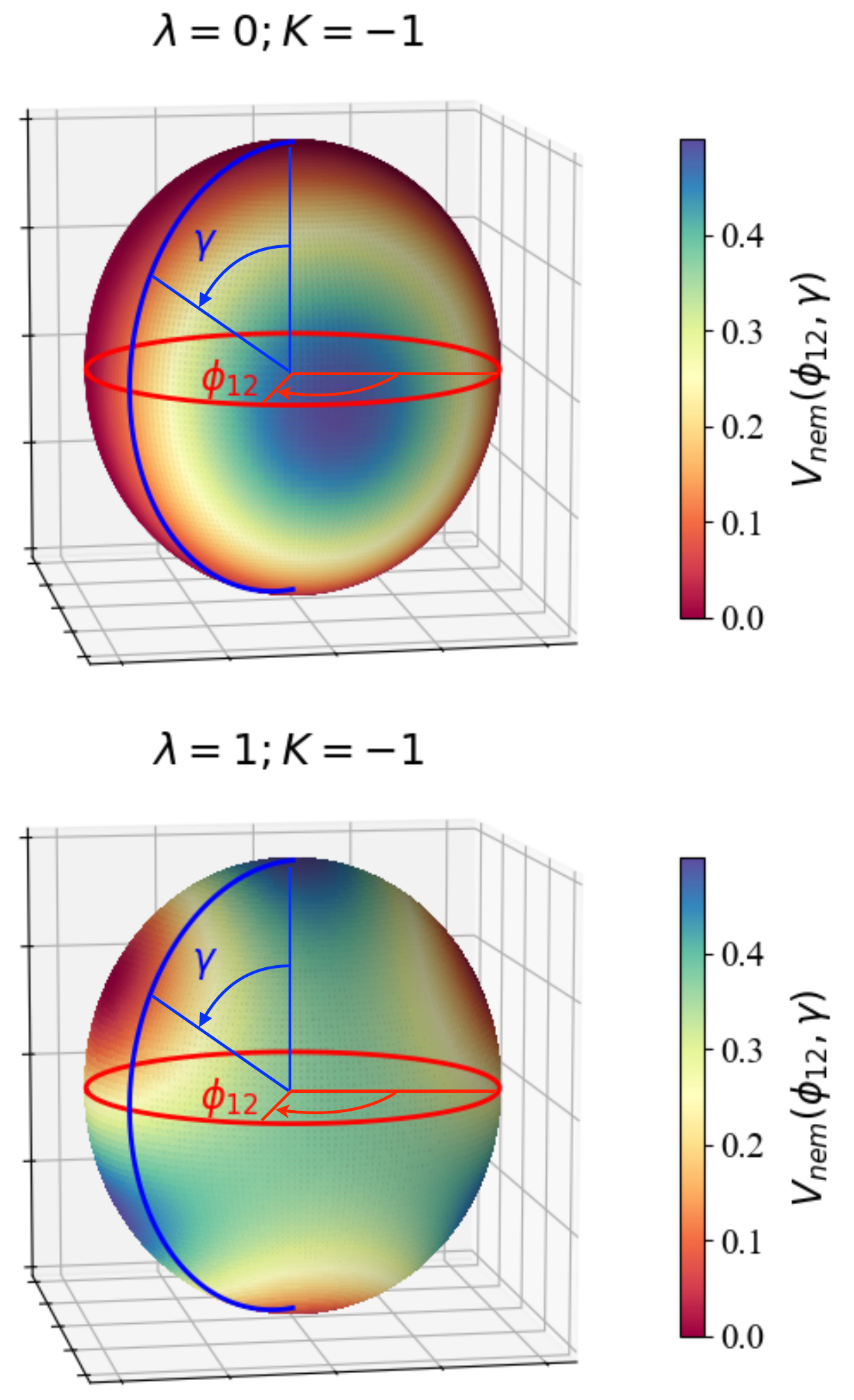}
    \caption{Nematic potential $V_{nem}(\phi_{12}, \gamma)$ in Eq.\eqref{potential_nem}, plotted around the unitary sphere defined by the azimuthal angle, $\phi_{12}$, and the polar angle, $\gamma$. Upper panel: For $K<0$ and $\lambda=0$, the system has an accidental degeneracy along $\gamma$. Lower panel: The degeneracy is lifted by higher-order potential terms. For $\lambda>0$ the potential define three equivalent minima.  }
    \label{fig:potential}
\end{figure}

\vspace{-0.5cm}
\section{Model}

We start with the simplest $U(1)\times Z_3$ Ginzburg-Landau model, which describes a two-component nematic superconducting order parameter $\vec{\Delta}=(\Delta_1, \Delta_2)$, with $\Delta_{1,2}=|\Delta_{1,2}|e^{i \phi_{1,2}}$: 

\beq
\begin{split}
f=& \frac{1}{2} \left( {\vec\nabla} \times {\vec A} \right)^2 +  \frac{1}{2} \sum_{\alpha=1,2}  \left|\left( \vec{\nabla} +e  {\vec A}\right)\Delta_\alpha\right|^2 + \\ &+V_{nem}(\Delta_1, \Delta_2),
\label{GLmodel}
\end{split}
\eeq

where we only include the simplest gradient terms, leaving out more general contributions \cite{zyuzin_nematic_2017,how_absence_2023,sigrist_phenomenological_1991}.  The leading potential terms, $V_{nem}(\Delta_1, \Delta_2)$, reads~\cite{sigrist_phenomenological_1991, chichinadze_nematic_2020}

\beq
\begin{split}
   V(\Delta_1, \Delta_2)&=  \alpha_1 \left( |\Delta_1|^2 +|\Delta_2|^2\right)  + \\ &+ \beta_1 ( |\Delta_1|^2  + |\Delta_2|^2)^2 + \beta_2\left| \Delta_1^2  + \Delta_2^2 \right|^2 + \\ & +\frac{\lambda}{2} \left[ (\Delta_1 -i\Delta_2)^3(\Delta^*_1 -i\Delta^*_2)^3 + c.c. \right].
\label{v1} 
\end{split}
\eeq

In Eq.~\eqref{v1}, the coefficient $\alpha_1 \propto (T - T_{c_0})$, where $T_{c_0}$ is the characteristic temperature at which the quadratic terms change sign. Stability requires $\beta_1 > 0$ and $\beta_1 + \beta_2 > 0$, while the sign of $\beta_2$ determines the symmetry of the superconducting order parameter.
The latter becomes evident when expressing the wavefunction on complex polar form,
$\Delta_\alpha =|\Delta_\alpha|e^{i\phi_\alpha}$, so that 
\begin{equation}
     \beta_2\left| \Delta_1^2  + \Delta_2^2 \right|^2 = 2\beta_2 |\Delta_1|^2|\Delta_2|^2\left[ \cos(2\phi_1 - 2\phi_2) -1\right].
\end{equation}

In what follows, we will assume a fixed total density $|\Delta_1|^2 + |\Delta_2|^2=1$, while retaining both relative density and phase fluctuations. {This approximation is valid in the regime where fluctuations of the total superconducting density occur at a much higher energy scale. This regime is favorable for increasing the temperature of the nematic $Z_3$ transition
since the energy cost of $Z_3$ domain walls can be made rather high. When the total
density can vary, large gradients at a domain wall can be diminished by locally reducing
the total density. As a result, the relative energy cost of a domain walls become cheaper, and thermal fluctuations can more easily restore $Z_3$ symmetry (see the analogous discussion for the 
$U(1) \times  Z_2$ case in Ref.~\cite{grinenko_state_2021}).} 

For convenience, here we define $K=2\beta_2$.
In this scenario, the potential nematic term, $V_{nem}(\Delta_1, \Delta_2)$, can be written as {a function of two angles $\phi_{12}=\phi_1 -\phi_2$ and $\gamma= \arctan{\frac{|\Delta_2|}{|\Delta_1|}}$ which parametrize the relative phase and relative amplitude between the two superconducting components}

\begin{equation}
\begin{split}
    &V_{nem}(\phi_{12}, \gamma)=-\frac{K}{2} \sum_{i}  \sin^2(\gamma)  \sin^2( \phi_{12})   + \\ &+ \lambda  \sum_i\left[ \cos(3\gamma) + 3 \cos(\gamma)  \sin^2(\gamma)\sin^2(\phi_{12}) \right].
    \end{split}
    \label{potential_nem}
\end{equation}

For $K>0$ and $\lambda=0$, the model describes a chiral superconducting ground state, where the intercomponent phase differences that minimize the free energy are $\phi_1 - \phi_2= \pm \pi/2$. For $K<0$ and $\lambda=0$, the system has a continuous  degeneracy { along $\gamma$, see upper panel of Fig. \ref{fig:potential}} which is lifted as higher-order terms, with coupling constant $\lambda$, are included, see lower panel in Fig.\ref{fig:potential}.  

The model in Eq.\eqref{GLmodel}, with $K<0$ and $\lambda\neq 0$, exhibits, along with the superconducting gauge symmetry $U(1)$, a $Z_3$ symmetry associated with three equivalent minima. Each minimum corresponds to a different, yet energetically degenerate, nematic order.
As shown in the lower panel of Fig.\ref{fig:potential}, for $\lambda>0$ the three minima are:
\begin{enumerate}
\item $\Delta_1 = \frac{\sqrt{3}}{2} e^{i\phi}; |\Delta_2| = \frac{1}{2} e^{i\phi};$
\item $\Delta_1 = \frac{\sqrt{3}}{2} e^{i\phi + \pi}; |\Delta_2| = \frac{1}{2} e^{i\phi};$
\item $\Delta_1 = 0; |\Delta_2| = e^{i\phi}.$
\end{enumerate}

By raising the temperature from a nematic superconducting ground state, a composite nematic fermionic order emerges if the two broken symmetries are restored at different temperatures such that $T_c^{U(1)} \leq T_c^{Z_3}$.
The possible emergence of these composite orders can be understood in terms of the competing proliferation of different kinds of topological defects \cite{svistunov_superfluid_2015}.
In nematic superconductors, the relevant topological defects include skyrmionic vortices, nematic domain walls, and fractional vortices \cite{zyuzin_nematic_2017,how2020half,wu2017majorana}.
A composite nematic phase, for instance, can be induced by the proliferation of single-quanta skyrmionic vortices. These defects restore the $U(1)$ gauge symmetry and make the system dissipative, without disrupting the coexisting $Z_3$ nematic order, leading to a scenario where $T_c^{Z_3} >T_c^{U(1)}$. 
Conversely, if the $Z_3$ domain walls--that do not carry a topological charge associated with the $U(1)$ gauge symmetry--proliferate first as the temperature increases, a charge-4e superconducting state may emerge as a composite order. 
In general, the separation of the two critical temperatures is hindered by the interactions between these topological defects. For example, the proliferation of skyrmionic vortices can trigger the proliferation of fractional vortices or $Z_3$ domain walls, and vice versa. In early works, assumptions of independent transitions led to incorrect conclusions about fluctuations in multicomponent superfluids \cite{korshunov1985two} (cf. with critical discussions in different formalisms \cite{How_Superfluid2024,yuan_absence_2024}).

Monte Carlo methods provide a powerful tool for determining whether the system exhibits multiple phase transitions.
In this work, we investigate the phase diagram of model Eq.\eqref{GLmodel} beyond the mean-field approximation by discretizing the model on a three-dimensional lattice and performing large-scale Monte Carlo simulations.  

\section{Monte Carlo Numerical Results}

Discretizing the model \eqref{GLmodel} we obtain:
\beq
\begin{split}
 &H = - \sum_{i,\mu=\hat{x}, \hat{y}, \hat{z}} \sum_{\alpha=1,2}| \Delta_{\alpha,i}| | \Delta_{\alpha,i +\mu} |  \cos{\chi_{\alpha, i}^\mu}+\\ & + \frac{1}{2} \sum_{\nu>\mu}   (F_{\mu\nu})^2 -\frac{K}{2} \sum_{i}  \sin^2(\gamma_i) \left[ \sin^2(\phi_{1,i} - \phi_{2,i}) \right] +\\& + \lambda  \sum_i\left[ \cos(3\gamma_i) + 3 \cos(\gamma_i)  \sin^2(\gamma_i)\sin^2(\phi_{1,i} - \phi_{2,i}) \right] , 
\end{split}
\label{hamiltonian}
\eeq
where $\chi_{\alpha, i}^\mu = \phi_{\alpha,i+\mu} -\phi_{\alpha,i} + eA_{i}^{\mu}$ is the gauge invariant superconducting phase and  $ F_{\mu\nu}= A_\mu(\mathbf{r}) + A_\nu(\mathbf{r+\mu}) - A_\mu(\mathbf{r+\nu}) - A_\nu(\mathbf{r})$ is the discrete form of the vector potential curl. 

Each Monte Carlo step consists of 50 local Metropolis-Hastings sweeps of all lattice fields followed by a parallel tempering swap of field configurations between neighboring temperatures. Each local sweep involves the two phase fields $\phi_1(\mathbf{r}), \phi_2(\mathbf{r}) \in [0, 2\pi )$, the two amplitude fields $|\Delta_{1}|, |\Delta_{2}|$, with the constraint $|\Delta_{1}(\mathbf{r})|^2 + |\Delta_{2}^2(\mathbf{r})|=1$, and the vector potential field $A_\mu(\mathbf{r})$. 
For most of the numerical simulations, we performed a total of $20^5$ Monte Carlo steps, with a transient time of maximum $20^4$ Monte Carlo steps. We use standard Bootstrap resampling methods to compute errobars.

\begin{center}
\textbf{The observables}
\end{center}

\begin{figure*}[t!]
    \centering
    \includegraphics[width= \linewidth]{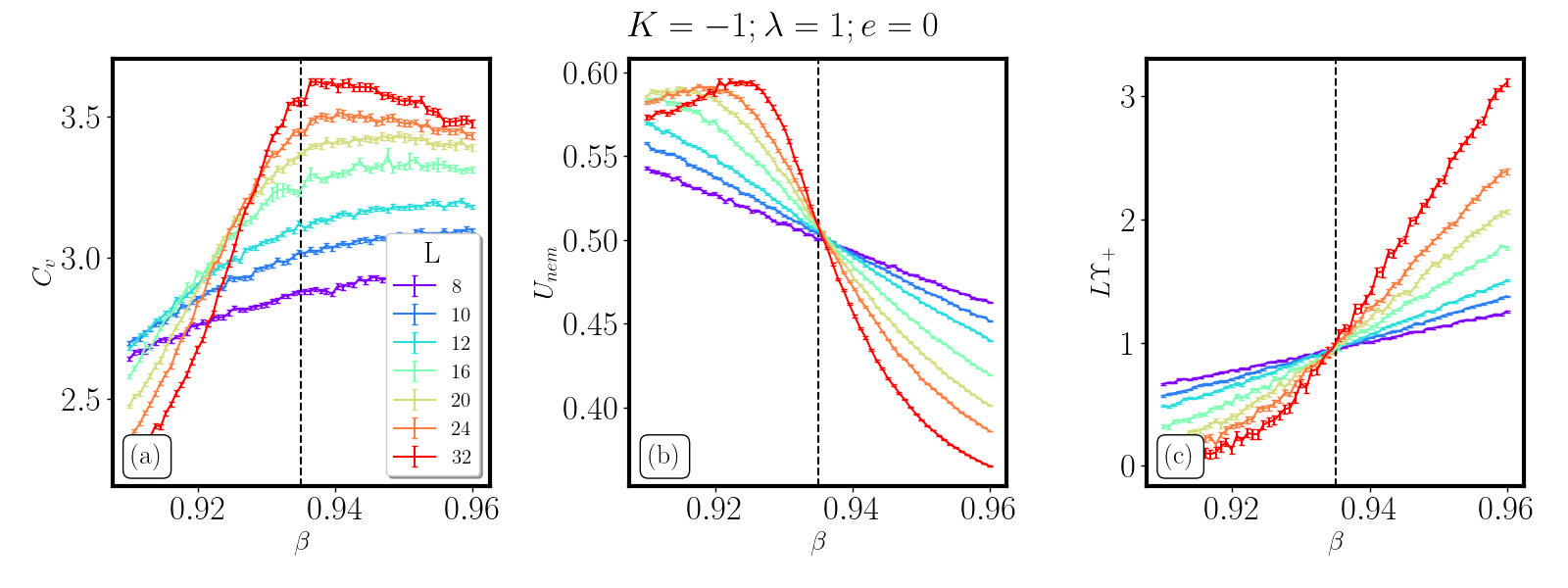}
    \caption{ Monte Carlo numerical results obtained for the case $K=-1$, $\lambda=1$ in the extreme type-II limit, i.e. $e=0$. The three panels show (a) the specific heat, $C_v$, (b) the nematic Binder cumulant, $U_{nem}$ and (c) the helicity modulus sum, $\Upsilon_+$, multiplied by the linear system size $L$ as a function of the inverse critical temperature $\beta=1/T$ and for different values of $L$. {Finite-size scaling analysis reveals the presence of a single phase transition: the specific heat exhibits a single peak, and the crossing points of $U_{\rm nem}$ and $L \Upsilon_+$ converge in the thermodynamic limit, indicating $\beta_c^{Z_3} = \beta_c^{U(1)}$. More details on the finite-size scaling analysis and the determination of $\beta_c$ can be found in Appendix A. In all panels, the dashed vertical black line indicates the extrapolated inverse critical temperature $\beta_c$.} }
    \label{fig_e_0}
\end{figure*}

Initially, we consider the case $e=0$, with no coupling to the vector potential. In this scenario, we identify the $U(1)$ phase transition, associated with a finite superconducting order parameter, by computing the helicity modulus sum, $\Upsilon^{+}$, which corresponds to the superfluid stiffness. This observable {measures the superconducting phase coherence by probing its response to an infinitesimal twist. Indeed, it} is defined as the linear response to an infinitesimally small twist of the superconducting phase along a given direction $\mu= \hat{x}, \hat{y}, \hat{z} $.
\begin{equation}
    \begin{pmatrix} \phi'_1(\mathbf{r}) \\  \phi'_2(\mathbf{r})  \end{pmatrix}
= \begin{pmatrix} \phi_1(\mathbf{r}) \\  \phi_2(\mathbf{r})  \end{pmatrix} 
+  \begin{pmatrix} {\delta}_\mu \cdot \mathbf{r} \\  {\delta}_\mu \cdot \mathbf{r}  \end{pmatrix} .
\end{equation}
The helicity modulus sum along $\mu$ is thus defined as
    \beq 
    \Upsilon^{\mu}_{+} = \frac{1}{L^3} \frac{\partial^2 F(\{\phi'_i\}) }{\partial \delta_{\mu}^2}\Bigr|_{\delta_{\mu}=0}= \Upsilon^{\mu}_{1} + \Upsilon^{\mu}_{2}+ 2 \Upsilon^{\mu}_{12},\eeq
where
\begin{equation}
\label{Helicity1}
\begin{split}
 \Upsilon^\mu_{i}= \frac{1}{L^3} \Big[ \Big\langle \frac{\partial^2 H}{ \partial \delta_{\mu,i}^2}
  \Big\rangle   -\beta \Big\langle \left(  \frac{\partial H}{ \partial \delta_{\mu,i}} - \langle \frac{\partial H}{ \partial \delta_{\mu,i}} \rangle \right)^2  \Big\rangle  \Big]_{\delta=0}, 
  \end{split}
 \end{equation}
 \begin{equation}
  \label{Helicity2}
 \begin{split}
 \Upsilon^\mu_{12}=  \frac{-\beta}{L^3} \Big[ & \Big\langle \frac{\partial^2 H}{ \partial \delta_{\mu,1} \partial \delta_{\mu,2} } \Big\rangle  -\langle \frac{\partial H}{ \partial \delta_{\mu,1}}\rangle \Big\langle \frac{\partial H}{ \partial \delta_{\mu,2}} \rangle   \Big\rangle \Big]_{\delta=0}.
 \end{split}
\end{equation}
Here, the brackets $\langle \dots \rangle$ imply thermal averages obtained via Monte Carlo simulations. For the model \eqref{hamiltonian}, $\Upsilon^\mu_{12} =0$ so that $\Upsilon^{\mu}_{+} =\Upsilon^{\mu}_{1} + \Upsilon^{\mu}_{2} $. In this work, we rely on the helicity modulus sum along $\mu=\hat{x}$, and for brevity, we denote $\Upsilon_{+}  \equiv   \Upsilon^x_{+}$. {Since $L\Upsilon_{+}$ is universal at the $U(1)$ critical point, we locate the superconducting transition by looking at the crossing point of $L\Upsilon_{+}$ for different system sizes. }

When $e\not=0$, and the gauge-field coupling is present, {the superfluid stiffness defined in Eq.~\eqref{Helicity1} is no longer an indicator of the presence or absence or the ordering in $U(1)$ sector. The phase sum $\phi_1 + \phi_2$ is not gauge-invariant, since any applied twist can be compensated by the vector potential $\vec{A}$ to which it couples. In this case,} the onset of superconductivity is determined by measuring the Meissner effects via the so-called dual stiffness~\cite{motrunich_comparative_2008, grinenko_state_2021, maccari_effects_2022}, defined as 
\beq
\rho^{\mu}(\mathbf{q})= \Big\langle  \frac{| \sum_{\mathbf{r}, \nu. \lambda} \epsilon_{\mu, \nu, \lambda} \Delta_{\nu} A_{\lambda}(\mathbf{r}) e^{i \mathbf{q \cdot r}}|^2}{(2 \pi)^2 L^3}   \Big\rangle.
\eeq
The dual stiffness vanishes in the superconducting phase, and becomes finite in the normal state, indicating the loss of diamagnetism. Here, we compute $\rho^{\mu}(\mathbf{q})$ along the $z$-direction for a small wave vector in the $x$-direction, $\mathbf{q}^x_{min}=(2\pi/L, 0,0)$, i.e. $\rho^z(\mathbf{q}^x_{min})$, which in the following we denote simply as $\rho$.
At the $U(1)$ critical point, similar to $\Upsilon_+$, the dual stiffness $\rho$ scales as $1/L$. Thus, for $e\neq 0$, the inverse critical temperature $\beta_c^{U(1)}$ can be located from the crossing points of $L\rho$ for different linear sizes $L$. 

To identify the $Z_3$ phase transition, associated with a finite nematic order parameter, we compute the local nematic order parameter: 

\begin{eqnarray}
\vec{N}_{i} &=&\left( N_{i}^x,  N_{i}^y, N_{i}^z \right) \\
 N_{i}^x &=&\Delta_{1,i} \Delta_{2,i}^* + \Delta_{1,i}^* \Delta_{2,i} \\  
 N_{i}^y &=&\Delta_{1,i} \Delta_{2,i}^* - \Delta_{1,i}^* \Delta_{2,i}\, \\
N_{i}^z&=& |\Delta_{1,i}|^2 - |\Delta_{2,i}|^2.
\end{eqnarray}

The nematic order parameter and the associated Binder cumulant are then defined respectively as
\beq
O_{\text{Nem}} = \frac{1}{L^3}\big[ \big(\sum_i N_{i}^x\big)^2  + \big(\sum_i N_{i}^y\big)^2  + \big(\sum_i N_{i}^z\big)^2 \big]^{1/2}
\label{nematic_mag}
\eeq

\beq
  U_{\text{nem}}= \langle O_{\text{Nem}} ^4 \rangle /\langle O_{\text{Nem}}^2\rangle^2.
  \label{nematic_binder}
\eeq
We identify the $Z_3$ critical temperature from the finite-size crossing points of $U_{\text{nem}}$ which is {is sensitive to changes in the probability distribution of the nematic order parameter and it is} expected to be universal at the critical point~\cite{binder_critical_1981}. 

Finally, we also compute the total energy of the system and the specific heat 
\begin{equation}
    C_v = \frac{\beta^2}{L^3}\left( \langle H^2\rangle - \langle H\rangle^2 \right).
    \label{cv}
\end{equation}


\begin{figure}[t!]
    \centering
    \includegraphics[width= \linewidth]{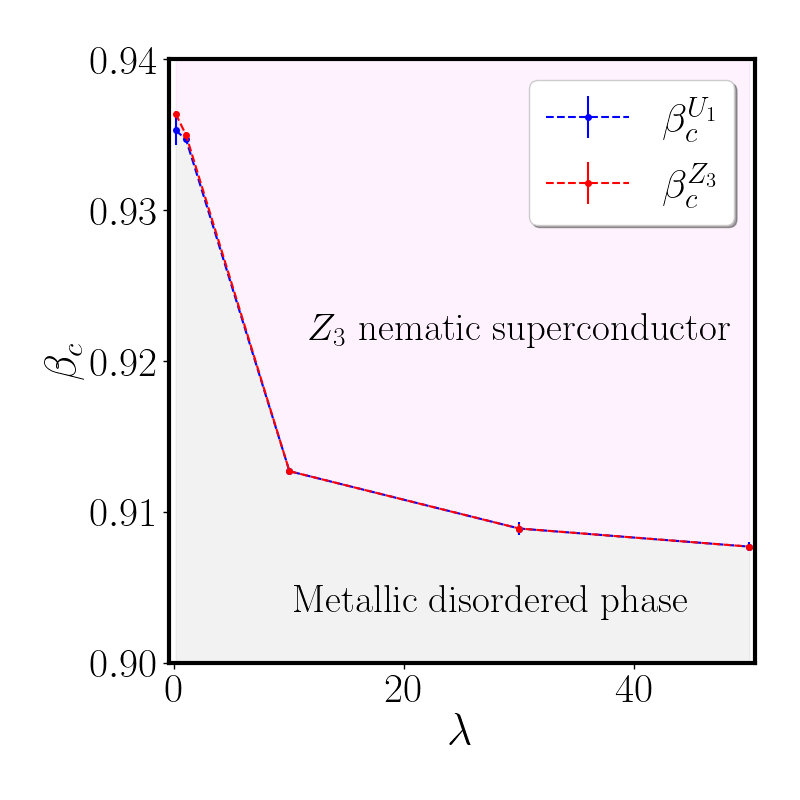}
    \caption{Phase diagram as a function of the nematic coupling $\lambda$ at fixed value of $K=-1$, and in the zero gauge-field limit.  The phase diagram reveals that the model at $e=0$ does not show a vestigial phase. For all the values of $\lambda$ investigated the two critical temperatures coincide.  In the phase diagram, where not visible, the error bars of the data points are smaller than the marker size. For details on the extraction of critical points, see the Supplemental Material.}
    \label{fig_phasediag_0}
\end{figure}

\vspace{-0.5cm}
\subsection{Phase diagram in the zero gauge-field limit}

We begin our investigation by considering the limit of vanishing gauge-field coupling, i.e., $e = 0$, which corresponds to a neutral system or to an extreme type II superconducting regime, characterized by a divergent London penetration depth $\lambda_L \propto 1/e \to \infty$.

In the limit $\lambda=K=0$, the model is $SU(2)$ symmetric and shows a single phase transition in the zero gauge-field limit~\cite{kuklov_deconfined_2008,herland_phase_2013, weston_composite_2021}. Finite values of $\lambda$ reduce the symmetry of the system from $SU(2)$ to $U(1)\times Z_3$. 
This leads to a reduced configuration space for fluctuations associated with the neutral mode, resulting in an increased critical temperature for the $Z_3$ nematic order as a function of $\lambda$. Consequently, as $\lambda$ increases, the only possible separation of the two critical temperatures is such that $T_c^{Z_3} > T_c^{U(1)}$.

We consider the model in Eq.\eqref{hamiltonian} for various values of $\lambda$ at a fixed $K = -1$. The numerical results for the case $\lambda = 1$, $K = -1$, and $e = 0$ are summarized in Fig. \ref{fig_e_0}. Finite-size scaling analysis {of the specific heat, Fig. \ref{fig_e_0}(a), the helicity-modulus sum, Fig. \ref{fig_e_0}(b), and the nematic Binder cumulant, Fig. \ref{fig_e_0}(c),} indicates that the system undergoes a single phase transition from a nematic superconducting ground state to a disordered metallic phase{, see Appendix A for more details on the extrapolation of the critical points to the thermodynamic limit}. 
The single-transition scenario appears to hold throughout the range of $\lambda$ values examined.  As shown in the phase diagram in Fig. \ref{fig_phasediag_0}, the two critical temperatures remain indistinguishable-- within error bars-- for all the values of $\lambda$ investigated, including very large values of the coupling constant. 
As evident in Fig. \ref{fig_phasediag_0}, in the limit $\lambda \to \infty$, the critical temperatures saturate due to the discrete lattice, which imposes a minimal size for the $Z_3$ domain walls separating different nematic configurations. 
Taken together, our results suggest that for $e = 0$, the model in Eq.\eqref{hamiltonian} does not exhibit a resolvable separation between the two critical temperatures and thus does not support the emergence of a composite ordered phase, in accordance with the findings of Ref.~\cite{how_absence_2023}. Further details on the finite-size scaling analysis used to extract the critical points are provided in Appendix A.

\begin{figure*}[t!]
    \centering
    \includegraphics[width=\linewidth]{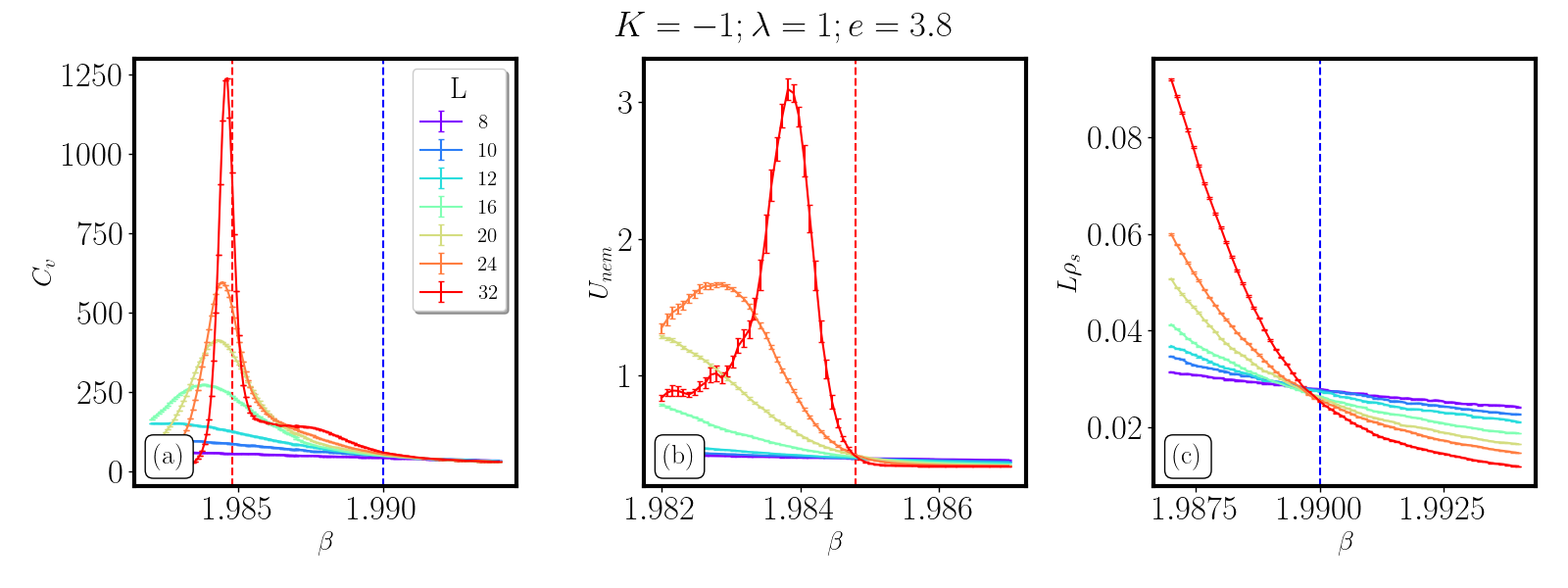}
    \caption{Monte Carlo numerical results obtained for the case $K=-1$, $\lambda=1$ and $e=3.8$. The three panels show (a) the specific heat, $C_v$, (b) the nematic Binder cumulant, $U_{nem}$ and (c) the dual stiffness, $\rho_s$, multiplied by the linear system size $L$, as a function of the inverse critical temperature $\beta=1/T$ and for different values of $L$. 
    {These numerical results demonstrate the existence of two distinct phase transitions, with $\beta_c^{Z_3} < \beta_c^{U(1)}$, as determined from the finite-size scaling analysis of $U_{\rm nem}$ and $L\rho_s$, respectively. This separation signals the emergence of an intermediate nematic nonsuperconducting phase in the regime $\beta_c^{Z_3}<\beta<\beta_c^{U(1)}$. Consistently, the specific heat exhibits two anomalies associated with the $Z_3$ and $U(1)$ critical points. While the $Z_3$ transition produces a sharp peak, the $U(1)$ anomaly is only weakly visible at these system sizes due to finite-size effects; it becomes slightly more pronounced for larger values of $e$, where the two transitions are more clearly separated (see Fig.~\ref{fig:e_5} in Appendix A). The dashed vertical red and blue lines mark the critical temperatures obtained from the crossing points of $U_{\rm nem}$ and $L\rho_s$, respectively. Further details on the finite-size scaling analysis can be found in Appendix~A.}
    }
    \label{fig_e_3.8}
\end{figure*}
\vspace{-0.2cm}
\subsection{Phase diagram as a function of the gauge-field coupling}

We now consider the different model with finite gauge field coupling, i.e. $e\neq0$. 
For convenience, we rewrite the free energy density Eq.\eqref{GLmodel} as
\beq
\begin{split}
&f=  \frac{1}{2 \rho^2} \left[ |\Delta_1|^2 \vec{\nabla} \phi_1 +  |\Delta_2|^2 \vec{\nabla} \phi_2 + e\rho^2 \vec{A} \right]^2  \\ +&\frac{|\Delta_1|^2|\Delta_2|^2}{2 \rho^2} \left[ \vec{\nabla} (\phi_1 - \phi_2)  \right]^2  + \frac{1}{2}\left[ (\vec{\nabla} |\Delta_1|)^2 + (\vec{\nabla} |\Delta_2| )^2\right]  \\  +&V_{nem}(\Delta_1, \Delta_2) + \frac{1}{2}(\vec{\nabla} \times \vec{A})^2,
\end{split}
\label{free_energy_gauge}
\eeq
with $\rho^2= |\Delta_1|^2 +|\Delta_2|^2 =1$. 
The electromagnetic field in Eq.\eqref{free_energy_gauge} couples to the co-flow of the two components, and thereby reduces the energy cost of skyrmionic vortices that carry integer flux \cite{zyuzin_nematic_2017}. 
Consequently, gauge field coupling drives fluctuations in the $U(1)$ sector and will eventually lead to a separation of the phase transitions so that $T_c^{Z_3} > T_c^{U(1)}$, provided that $e$ is sufficiently large. 
This will give rise to a non-superconducting composite order state that breaks the $Z_3$ nematic symmetry.

Our simulations confirm this alternative scenario. For gauge couplings $e < 3.5$, within the accuracy of our calculations, the system exhibits a single phase transition from a nematic superconductor to a metallic, disordered phase. However, upon further increasing the gauge coupling, we observe a clear splitting of the two phase transitions.
In Fig.~\ref{fig_e_3.8}, we show the specific heat, the nematic Binder cumulant, and the dual stiffness as functions of inverse temperature $\beta$ for $e = 3.8$. Here, the two inverse critical temperatures are distinctly separated, with $\beta_c^{Z_3} < \beta_c^{U(1)}$, and the specific heat displays two well-defined peaks corresponding to the $Z_3$ nematic and the $U(1)$ superconducting transitions.
The phase diagram, as a function of the electric charge $e$, is shown in Fig. \ref{fig:phase_diagram}. 

As discussed in Appendix A, our numerical simulations indicate that near the bicritical point where the two critical temperatures separate, the transition acquires a more pronounced first-order character{, as revealed by the emergence of a bimodal energy distribution and by the convergence of the two critical temperatures in the finite-size scaling analysis.}.  

\begin{figure}[b!]
    \centering
    \includegraphics[width=\linewidth]{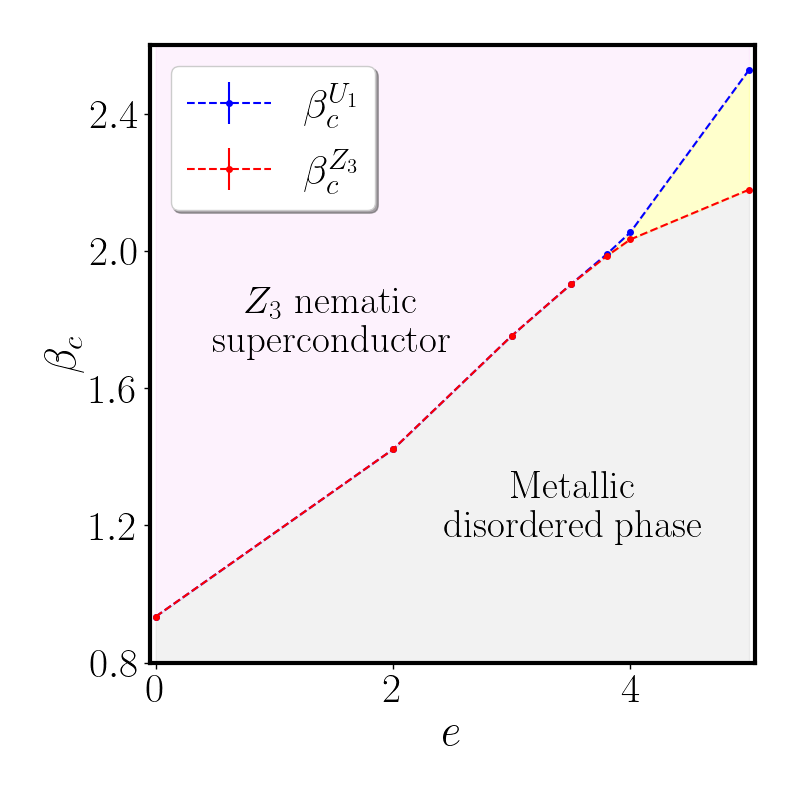}
    \caption{Phase diagram as a function of the electric charge $e$ for the free energy with parameters $\lambda=|K|=1$. Where not visible, the error bars of the data points are smaller than the marker size. This phase diagram reveals that a composite nematic phase can be resolved at large enough values of the electric charge $e>3.5$. }
    \label{fig:phase_diagram}
\end{figure}
\vspace{-0.2cm}

\subsection{Electromagnetic stabilization of the nematic composite order} 
 
Our numerical results have revealed that, in three spatial dimensions, a nematic composite order is observable only for large values of the intercomponent gauge-field coupling. 
However, in two spatial dimensions, the conditions for realizing such a phase are less restrictive.

In two dimensions, the inclusion of a gauge-field coupling also alters the energetics of topological defects: it renders the energy cost of creating defects that carry an integer number of flux quanta finite. In nematic systems, these defects typically take the form of skyrmionic vortices, which also possess finite energy~\cite{zyuzin_nematic_2017,speight2023magnetic}. As a result, the arguments presented in Ref.~\cite{babaev2002phase} become directly applicable: while superconducting order is destroyed at any finite temperature due to thermal fluctuations in the thermodynamic limit, the discrete $Z_3$ nematic order can remain stable and persist to finite temperatures.


In three dimensions, the situation differs markedly. Unlike in two dimensions, superconductivity can survive at finite temperatures because of the finite line tension of vortices. In a three-dimensional nematic system, skyrmionic vortices have a finite energy per unit length, creating a significant energetic obstacle to their proliferation. This results in the difficulty of stabilizing a composite nematic order in simple three-dimensional nematic models without invoking additional mechanisms.

An alternative way to enlarge the parameter space to stabilize a nematic composite order is by suppressing superconductivity externally, for instance, by applying an external magnetic field, inducing a dilute superconducting vortex lattice.
Such a vortex lattice can be melted with relatively weak thermal fluctuations, thereby restoring the gauge symmetry, while leaving the nematic symmetry broken. Thus, we propose that a composite nematic state in three-dimensional superconductors could be catalyzed through vortex lattice melting induced by an external magnetic field. In addition to the standard signatures of nematic order, one should observe a distinct specific heat anomaly in the vortex liquid phase above the latent heat peak associated with the melting transition.

\section{Conclusions}
In summary, we have conducted large-scale Monte Carlo simulations to assess the propensity for composite/vestigial order in three-dimensional nematic superconductors. In the absence of gauge-field coupling, for the commonly used model, our results show a single phase transition from a nematic superconducting state to a disordered phase, with no evidence of an intermediate composite order/vestigial state. Our simulations are consistent with analytical calculations in \cite{how_absence_2023}.

When gauge-field coupling is included, we observe that a vestigial nematic order with broken $Z_3$ symmetry--emerges above the superconducting transition. Yet, it is only resolvable in our model for very strong gauge coupling ($e > 3.5$).
While this serves as a proof of principle, it also suggests that the realization of such phases via a gauge-field coupling in materials similar to Bi$_2$Se$_3$-based candidates may require suppressing the superconducting transition by applying an external magnetic field--following an argument similar to that in Refs.~\cite{babaev_superconductor_2004,smorgrav_observation_2005}--which leads to the melting of a dilute lattice of nematic skyrmionic vortices.
Alternatively, these phases could be stabilized in the absence of an external field through additional intercomponent interactions, such as mixed gradient terms, arising, for example, from strong correlations, similar to those considered in Refs.~\cite{grinenko_state_2021, maccari_effects_2022}.

\vspace{-0.2cm}

\section*{Acknowledgments}
\vspace{-0.2cm}
The computations were enabled by resources provided by the National Academic Infrastructure for Supercomputing in Sweden (NAISS), partially funded by the Swedish Research Council through grant agreement no. 2022-06725. 
EB was supported by the Swedish Research Council Grants  2022-04763, by Olle Engkvists Stiftelse,  and partially by the Wallenberg Initiative Materials Science
for Sustainability (WISE) funded by the Knut and Alice Wallenberg
Foundation.
JC was supported by the Swedish Research Council (VR) through grant 2018-03882.
 IM acknowledges financial support by the Swiss National Science Foundation (SNSF) via the SNSF postdoctoral Grant No. TMPFP2\_217204.
JC and IM were supported by the Carl Trygger foundation through Grant No. CTS 20:75.
JC and EB were supported by a project grant from Knut och Alice Wallenbergs Stiftelse.

\vspace{-0.2cm}
\section*{Appendix A: Assessment of the critical points}

To assess the critical temperatures $U(1)$ and $Z_3$, we extract the finite-size crossing points of the relevant observables and extrapolate their thermodynamic values. 
Here, we use as an ansatz the finite-size scaling behavior appropriate to the nature of the
corresponding phase transition.

The superconducting ${U(1)}$ phase transition, at $e=0$, belongs to the same universality class as the three-dimensional XY model, whose inverse critical temperature finite-size scaling reads~\cite{Campostrini_3DXY_2001} 
    \beq
    \beta_c^{U(1)}(L)= \beta_c^{U(1)}(\infty) + b L^{1/\nu}. 
    \label{fitU1}
    \eeq
We used the same ansatz for case $e\neq0$.

\begin{figure}[b!]
    \centering
    \includegraphics[width=0.85\linewidth]{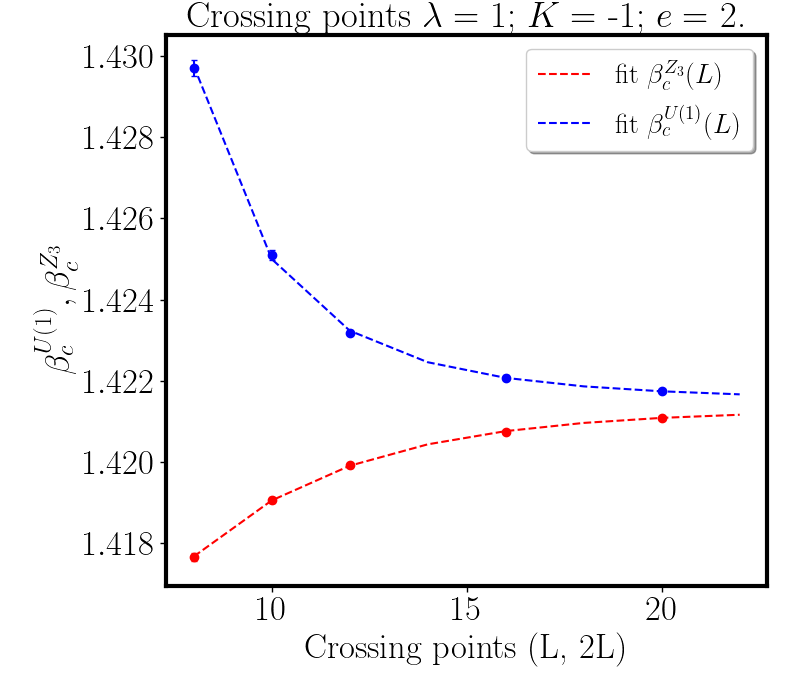}
    \caption{Finite-size scaling of the crossing points $\beta_c^{Z_3}$ and $\beta_c^{U(1)}$ extracted from the nematic Binder cumulant and the dual stiffness for the case $e=2.0$, $\lambda=1$, and $K=-1$. The two fitting functions give $\beta_c^{U(1)}=\beta_c^{Z_3}=1.421$. }
    \label{crossing_e2}
\end{figure}

On the other hand, the nematic $Z_3$ phase transition is expected to display the same
symmetry and finite-size behavior as the three-state Potts model, which in three spatial
dimensions exhibits a weak first-order phase transition~\cite{wu_potts_1982, janke_three-dimensional_1997}. Thus, we use the corresponding finite-size scaling~\cite{janke_three-dimensional_1997}:
\beq
    \beta_c^{Z_3}(L)= \beta_c^{Z_3}(\infty) + b e^{-\nu L}. 
        \label{fitZ3}
\eeq
Let us highlight that these scaling functions do not have to necessarily hold for our model Eq.\ref{hamiltonian}, since in a wide range of the parameter space the system undergoes a single $U(1) \times Z_3$ phase transition, which is expected to be first order.
An example of the extrapolation of the two critical temperatures through these two scaling functions is shown in Fig.\ref{crossing_e2}.
\begin{figure}[h!]
    \centering
    \includegraphics[width=0.85\linewidth]{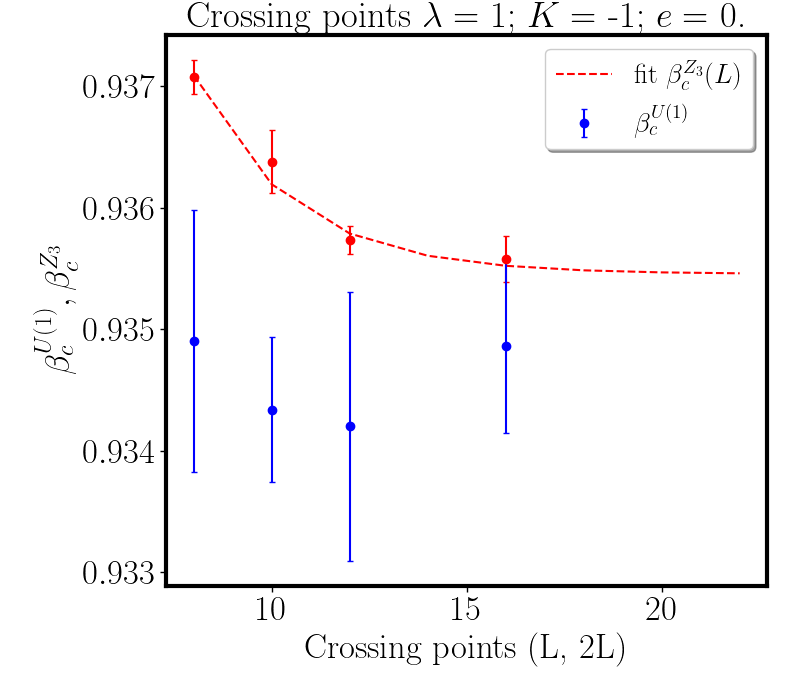}
    \caption{Finite-size scaling of the crossing points $\beta_c^{Z_3}$ and $\beta_c^{U(1)}$ extracted from the nematic Binder cumulant and the helicity modulus sum for the case $e=0$, $\lambda=1$, and $K=-1$. }
    \label{crossing_e0}
\end{figure}
\begin{figure}[b!]
    \centering
    \includegraphics[width=0.85\linewidth]{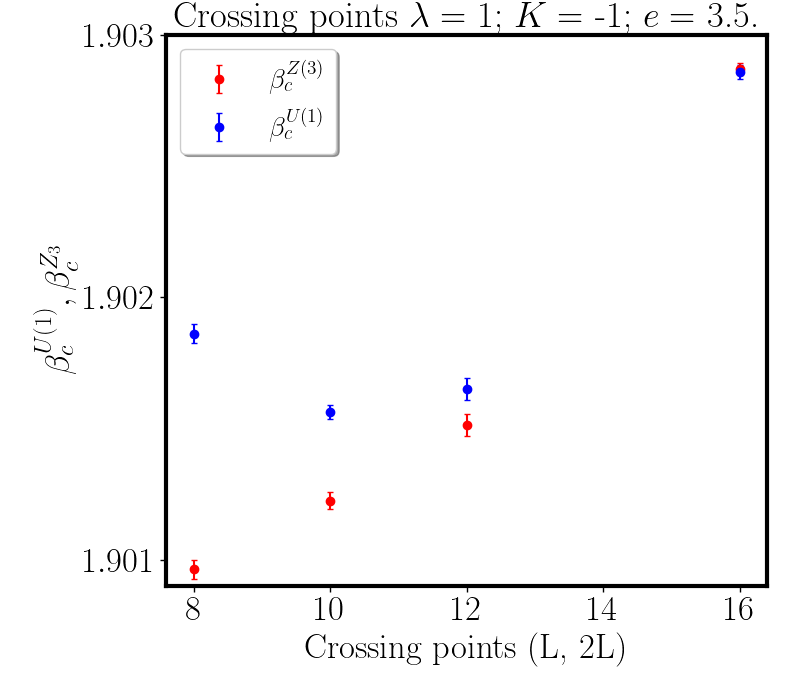}
    \caption{Finite-size scaling of the crossing points $\beta_c^{Z_3}$ and $\beta_c^{U(1)}$ extracted from the nematic Binder cumulant and the dual stiffness for the case $e=3.5$, $\lambda=1$, and $K=-1$. The strong drift of the crossing points is another indication of the proximity to a critical point where the two phase transitions split apart. }
    \label{crossing_e3.5}
\end{figure}
\begin{figure*}[t!]
    \centering
    \includegraphics[width=0.9\linewidth]{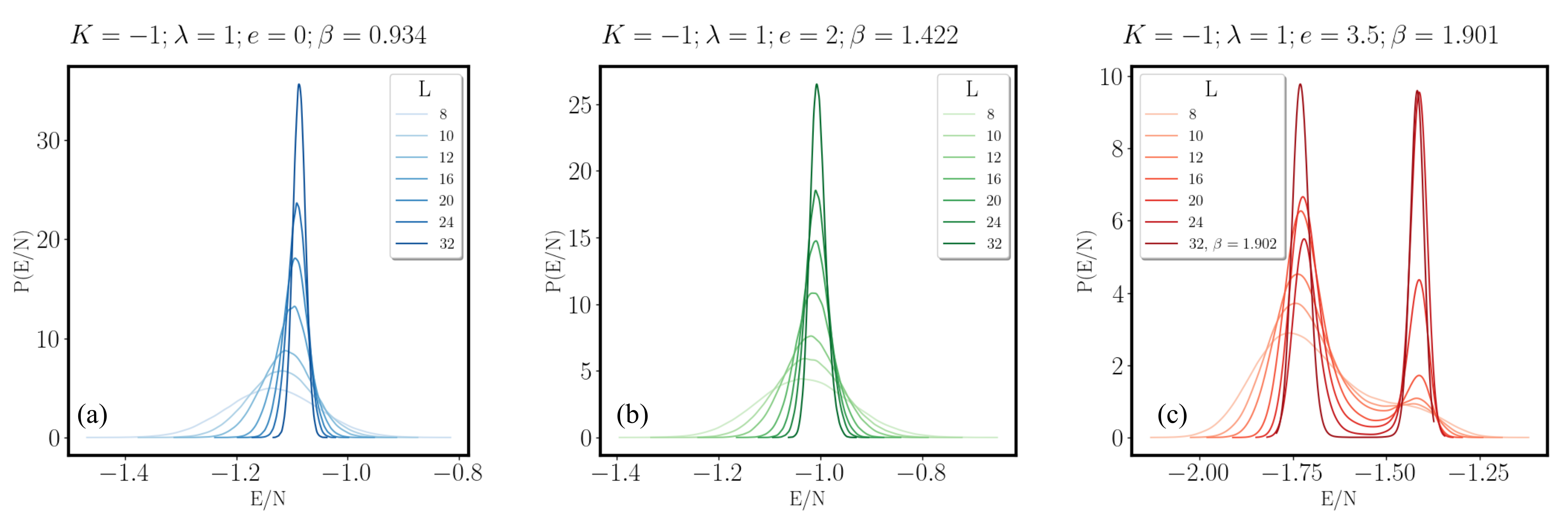}
    \caption{Probability distribution of the energy per lattice site, $E/N$, for different values of the linear size at the critical point and with system parameters $\lambda=1$, $K=-1$, and: (a) $e=0$; (b) $e=2.0$; (c) $e=3.5$. By increasing the value of the gauge coupling and approaching the critical value at which the two phase transitions split apart, we identify a possible tricritcal point where a second order phase transition, for small values of $e$, turns into a first order phase transition at $e>2$. The assessment of this tricritical point is beyond the scope of the present work. }
    \label{double_peaks}
\end{figure*}

For the cases where we could not properly fit the finite-size crossing points via Eqs.\eqref{fitU1}-\eqref{fitZ3}, we take the crossing point of the largest finite sizes, for example, the one of $(L, 2L)=(16, 32)$ in Fig.\ref{crossing_e0} and Fig.\ref{crossing_e3.5}.
Finally, it is worth noticing that by increasing the value of the gauge coupling $e$, and approaching the point at which the two phase transitions split apart, the transition acquires a more pronounced first-order character which manifests itself in a double-peak feature of the energy distribution, see panel (c) of Fig.\ref{double_peaks}. 
This is similar to other multicomponent models considered in other contexts \cite{kuklov_deconfined_2006,kuklov_deconfined_2008}. At the same time, by approaching the bicritical point, the finite-size effects also become much more pronounced, as is visible from the finite-size analysis of the crossing points shown in Fig.\ref{crossing_e3.5}.
\vspace{-0.5cm}
\section*{Appendix B: Data for $e=5$ with largest critical temperatures separation }
\begin{figure*}[t!]
    \includegraphics[width=0.9\linewidth]{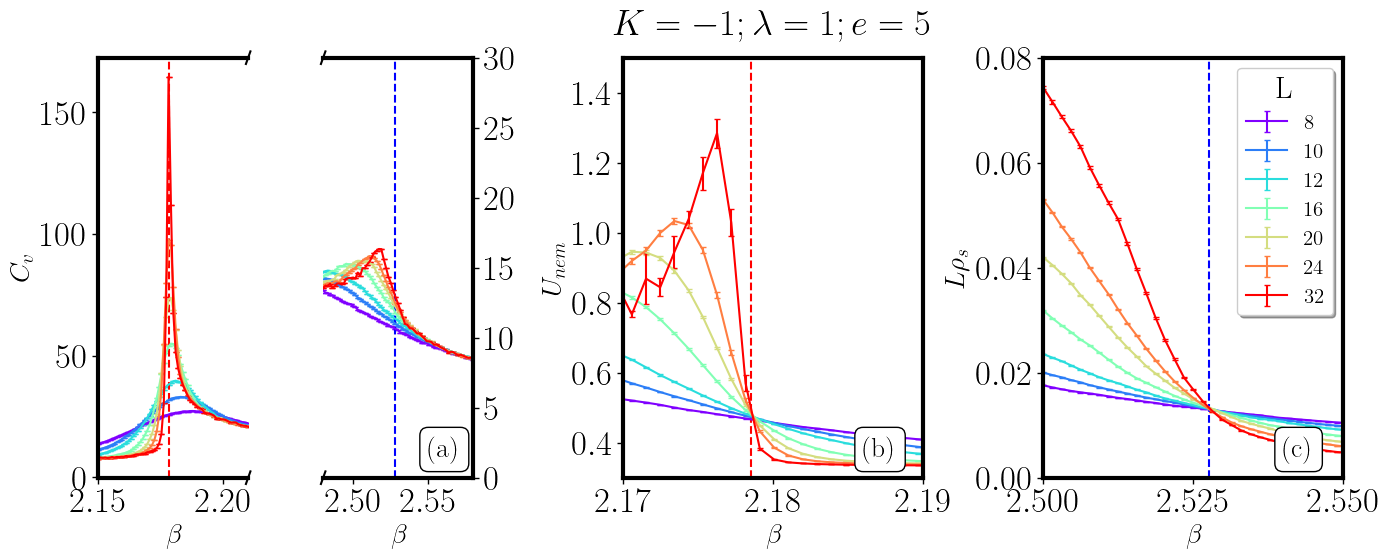}
    \caption{Monte Carlo numerical results obtained for the case $K=-1$, $\lambda=1$, and $e=5$. The three panels show (a) the specific heat, $C_v$, (b) the nematic Binder cumulant, $U_{\rm nem}$, and (c) the dual stiffness, $\rho_s$, multiplied by the linear system size $L$, as functions of the inverse temperature $\beta=1/T$ for different values of $L$. In this regime, the two phase transitions are more widely separated, leading to a clearly visible $U(1)$ peak in the specific heat in addition to the sharp $Z_3$ peak. The dashed vertical red and blue lines indicate the critical points $\beta_c^{Z_3}$ and $\beta_c^{U(1)}$, extracted from the finite-size scaling analysis of $U_{\rm nem}$ and $L\rho_s$, respectively.}
    \label{fig:e_5}
\end{figure*}

\bibliography{mylib}

\end{document}